\documentclass[prb,preprint]{revtex4}
\usepackage{epsfig}

\usepackage{amsmath}

\begin{document}

\title
{Transmission through a non-overlapping well adjacent to a finite barrier}

\author{Zafar Ahmed}
\email{zahmed@barc.gov.in}
\affiliation{Nuclear Physics Division, Bhabha Atomic Research Centre,
Mumbai 400 085, India}


\begin{abstract}
We point out that a non-overlapping well (at negative energies) adjacent to a finite barrier (at positive energies)
is a simple potential which is generally missed out while discussing the one-dimensional potentials in the textbooks of quantum mechanics.
We show that these systems present interesting situations wherein transmitivity $(T_b(E))$ of a finite barrier  can be changed
both quantitatively and qualitatively by varying the depth or width of the well or by changing the distance 
between the well and the barrier. Using delta (thin) well near a delta (thin) barrier we show that the well induces
energy oscillations riding over $T_b(E)$ in the transmitivity $T(E)$ at both the energies below and above the barrier.
More generally we show that a thick well separated from a thick barrier also gives rise to energy oscillations in $T(E)$.
A well joining  a barrier discontinuously (a finite jump) reduces $T(E)$ (as compared to $T_b(E))$
over all energies. When the well and barrier are joined continuously, $T(E)$ increases and then decreases at energies 
below the barrier. At energy above the the barrier the changes are inappreciable. In these two cases  if we separate the well 
and the barrier by a distance, $T(E)$ again acquires oscillations. Paradoxically, it turns out that a distant well induces more 
energy oscillations in $T(E)$ than when it is near the barrier.
\end{abstract}

\pacs{PACS No.: 03.65.Ge, 03.65.Nk}
\maketitle
\section{Introduction}
In the textbooks of quantum mechanics the solution of Schr{\"o}dinger equation and the consequent results are illustrated through simple 
one-dimensional potentials. For discrete bound states  the square well\cite{1,2,3,4} and double wells\cite {2,3} are studied. 
Square well, square barrier and semi-infinite step potentials are used for studying continuous energy (scattering) states.\cite{2,3,4} 
A well with two side barriers is studied for understanding resonances and meta-stable states.\cite{2,3} An overlapping well adjacent 
to a finite barrier is a well known model  for discussing discrete complex energy Gamow-Seigert meta-stable states \cite{5} in alpha decay.

Students may wonder as to what happens if a non-overlapping well (at negative energies) is adjacent to a finite barrier (at
negative energies) (see Figs.~1). Perhaps for the want of an application this system has gone undiscussed, however, interesting queries 
do arise for this kind of potentials. One may wonder as to whether the well (at negative energies) can change (increase/decrease) 
the transmitivity of the barrier (at positive energies) quantitatively and significantly. One may like to know whether there 
can be qualitative changes in the transmitivity of the barrier $(T_b(E))$ due to the presence of the well in some class of cases.

In this article we would like to show that a well near a barrier can change the transmitivity of the barrier both quantitatively and qualitatively.
In fact a scattering potential well (vanishing at $x \rightarrow \pm \infty$) can give rise to a non-overlapping well adjacent to a finite barrier
(NWAFB) as 
\begin{equation}
V(x)=-v_w f(x+d) + v_b f(x),
\end{equation}
where $f(x)=e^{-x^2}, {\rm sech}^2x, e^{-x^4},....$ see Figs.~1(a). However in this case, a change in the depth of the well or its distance from the barrier
would also change the height of the barrier. Consequently, the effect of the well on the transmission property of the original barrier
can not come up explicitly. We, therefore, consider wells of zero-range or finite range. Else, if they are scattering wells of infinite range 
on one side they ought to be joined to the barrier continuously or dis-continuously. In the following we discuss the various possibilities 
for NWABF.
 
\section{Various models of non-overlapping well adjacent to a finite barrier}
We construct various models of NWAFB using three parameters $v_w,v_b>0$ and $d$. Here $v_w$ is the depth of the well, 
$v_b$ is height of the barrier and $d$ denotes the separation between the well and the barrier. In these models a change 
in $d$ does not change the depth of the well or the height of the barrier.

First let us consider both the well and the barrier of zero range.
Using the zero range Dirac delta potentials we construct a simple solvable model of NWAFB as
\begin{equation}
V^{\delta}(x)= -v_w \delta(x+d)+ v_b \delta(x). 
\end{equation}
Using finite range well, we construct a more general  model of NWAFB (see Figs.~1(b))
\begin{eqnarray}
&&V^F(x)= -v_w V_w(x+d+w_w/2), \quad -d-w_w \le x \le -d \nonumber \\ 
&&V^F(x)=0, \quad -d \le x \le 0, \nonumber \\
&&V^F(x)= v_b V_b(x), \quad  x \ge 0, 
\end{eqnarray}
where $V_w(x)$ may be chosen as constant (square or rectangular well), $(1-4x^2/w_w^2)$ (parabolic well), $(1-2|x|/w_w)$ (triangular well),
$e^{-x^2/w_w^2}$ (Gaussian well) or $e^{-|x|/w_w}$ (exponential well). 
It may be mentioned that in some cases $v_b$ may not represent the effective barrier height ($v_m=$maximum of $V_b(x)$).  
For instance in this article we shall be choosing $V_b(x)=v_b x e^{-x^2}$ where for $v_b=11.5$
we get $v_m \approx 5$.
 
Using asymptotically converging profiles $f(x)$ and $g(x)$, we construct two-parameter $(v_w,v_b)$ models of NWABF 
wherein a well of infinite range is juxtaposed to a barrier of infinite range continuously as (see solid curve in Figs.~1(c)) 
\begin{eqnarray}
&&V^{C}(x)= v_b g(x) , x>0 \nonumber \\
&&V^{C}(x)=v_w g(x), x \le 0, 
\end{eqnarray} 
and discontinuously as (see dashed curve in Figs.~1(c))
\begin{eqnarray}
&&V^{D}(x)=v_b f(x), x>0 \nonumber \\
&&V^{D}(x)=-v_w f(x) , x \le 0. 
\end{eqnarray} 
Here the functions $f(x)$ may be chosen as rectangular profile or as $e^{-x^2}$, $e^{-x^4}$, ${\rm sech}^2 x$...,
and $g(x)$ may be taken as $xe^{-x^2}$, $xe^{-x^4}$, $\tanh x~{\rm sech} x$,... . It may be mentioned that the finite range potential
like $V(|x|\ge w)=0, V(x<0)=v_w \sin (2\pi x/w), V(x>0)=v_b \sin (2\pi x/w)$ would rather be a NWAFB of type (3) with $d=0$ than of
the type (4).

Next we have to solve the Schr{\"o}dinger equation 
\begin{equation}
{d^2 \psi(x) \over dx^2}+{2m \over \hbar^2}(E-V(x)\psi(x)=0.
\end{equation}
for finding the transmitivity, $T(E)$, of the various potential models discussed above.
When the potentials are real and Hermitian the time reversal symmetry ensures that the transmitivity and reflectivity are independent of the direction 
of incidence of particle whether it is from left or right. Due to this symmetry, in transmission through NWAFB it does not matter whether the incident
particle sees the well or the barrier first.

\section{Delta potential model of NWAFB: (2)}
The zero range delta potential model of NWAFB is exactly solvable. We solve the Schr{\"o}dinger equation (6)
for this potential, $V^{\delta}(x)$ given in Eq.~(1) using just plane waves: $e^{\pm ikx}$ as usual. Let the direction of incidence
of the particle at the potential be from the left hand, we can write
\begin{eqnarray}
&&\psi(x)=A e^{ikx}+ B e^{-ikx}, \quad  -\infty < x \le -d \nonumber \\
&&\psi(x)=C e^{ikx}+ D e^{-ikx}, \quad -d < x < 0 \nonumber \\
&&\psi(x)=F e^{ikx}, \quad x \ge 0.
\end{eqnarray}
The wavefunction (7) has to be continuous at $x=-d$ and 0. However, due the point singularity at $x=-d,0$ in delta functions in Eq.~(2), there occurs 
a mis-match in the first derivative (see Problem no. 20 and 21 in Ref.\cite{4})  of the wavefunction we get
\begin{eqnarray}
&&A e^{-ikd} + B e^{ikd} = C e^{-ikd} + D e^{ikd}, \nonumber \\
&&ik[A e^{-ikd} - B e^{ikd}]-ik[C e^{-ikd} - D e^{ikd}] = -{2m \over \hbar^2} v_w [C e^{-ikd} + D e^{ikd}], \nonumber \\
&&C+D=F,\nonumber \\
&&ik[(C-D)-F]={2m \over \hbar^2} v_b F.
\end{eqnarray}
by eliminating $C,D$ and $F$ from Eq.~(8),  we get
\begin{eqnarray}
&& {B \over A}= {u_w (2ik+u_b)+\lambda^2(2ik+u_w) u_b \over (2ik-u_w)(2ik+u_b)+\lambda^2 u_w u_b}, \nonumber \\
&& {F \over A}={4k^2 \over (2ik-u_w)(2ik+u_b)+\lambda^2 u_w u_b}, \quad \lambda=e^{2ikd}, u_w={2m v_w \over \hbar^2}, u_b={2m v_b \over \hbar^2}.
\end{eqnarray}
These ratios give  us the reflectivity $R(E)=|{B \over A}|^2$ and the transmitivity $T(E)=|{F \over A}|^2$. When $v_w=v_b$
the numerator of $B/A$ in Eq.~(9) becomes $\cos ka$ which gives rise reflectivity zeros when $ka=(n+1/2)\pi$ these are the positions of transmission 
resonances with $T(E)=1.$ When either of $v_w$ and $v_b$ is zero, from Eq.~(9) we get (see Problem no. 21 in \cite{4})
\begin{equation}
T_b(E)= {E \over {m v^2 \over 2 \hbar^2}+E}=T_w(E), \quad v=v_w, v_b.
\end{equation}
This is a particular feature of the delta potential well or barrier that their transmission co-efficients are identical.
For all our calculations we choose $2m=\hbar^2=1$, so that energies and lengths are in arbitrary units.
In Figs.~2(a), both $T(E)$ and $T_b(E)$ are plotted as a function of energy, $E$, when $v_w=1, v_b=5, d=3$.
See the interesting energy-oscillations in solid curve that represent the transmitivity of the total potential
$V^{\delta}(x)$: a perturbed barrier. When compared with the transmitivity of the Dirac delta barrier (see the dotted curve) these 
energy oscillations in  $T(E)$ can be seen to be riding around $T_b(E)$ even at large energies ($E>>v_b$).
We find that the smaller values of $v_w$ (than 1) create only small excursions (ripples) around the smooth variation of $T_b(E)$.
 
The depth of the well $v_w$ governs the amplitude of these oscillations. 
In Figs.~2(c) see that the frequency of these energy-oscillations remain the same but their amplitudes are larger as $v_w$ is 
increased and made equal to 5$(=v_b)$. Compare Figs.~2(a) with Figs.~2(c) and Figs.~2(b) with Figs.~2(d) to appreciate the effect
of the increase in the depth of the well resulting in the increase of amplitude of oscillations.

We find that the frequency of these oscillations is governed by the value of $d$. Larger the value of $d$, more is the frequency of oscillations. 
Compare figs.~2(a) with  Figs.~2(b) and Figs.~2(c) with Figs.~2(d) to appreciate the effect of the increase in $d$. 
 
This simple and exactly solvable model of NWAFB suggests that a well near a barrier neither increases nor decreases the transmitivity
of the barrier. Most interestingly, it does both and hence energy oscillations in $T(E)$. Increase in the frequency 
of these oscillations due to increase in $d$ (perturbation moving away) is paradoxical.
 
The question arising here is whether energy oscillations in $T(E)$ is the essence of NWAFB of some type or a particular 
feature of extremely thin delta potentials making up $V^{\delta}(x)$ (2). We therefore need to study the other models given Eqs.~(1,3-5).
As the other models of NWAFB are not solvable analytically, in the following we discuss a numerical procedure to find $T(E)$.

\section{A numerical method for the calculation of transmitivity of a one dimensional potential} 
When the potentials vanish asymptotically one can calculate its transmission co-efficient by solving
the Schr{\" o}dinger equation numerically for scattering solutions.
We propose to solve Eq.~(6) using Runge-Kutta method\cite{6} of step by step integration (see Appendix). 
This method consists 
of solving two first order, linear, one dimensional coupled differential equations
\begin{equation}
{dy(x)\over dx}=f[x,y(x),z(x)], \quad {dz(x)\over dx}=g[x,y(x),z(x)], \quad y(0)=c_1, \quad z(0)=c_2.
\end{equation}
In this setting, we introduce $y(x)=\psi(x)$ and $z(x)= {d \psi(x) \over dx}$ 
and split the Schr{\"o}dinger equation in two first order coupled linear differential equations as 
\begin{eqnarray}  
{dy(x) \over dx}= z(x)\nonumber \\
{dz(x) \over dx}= -{2m \over \hbar^2}[E-V(x)]y(x).
\end{eqnarray}
The Schr{\" o}dinger equation which is a second order differential equation will have two linearly independent solutions
as $\psi_1(x)$ and $\psi_2(x)$. We start the numerical integration from $x=0$ using the two sets of initial values as
(see Problem no. 22 in Ref.\cite{4} and Ref.\cite{7})
\begin{equation}
\psi_1(0)=1,\psi_1^\prime(0)=0; \quad \psi_2(0)=0, \psi_2^\prime(0)=1,
\end{equation}
such that the Wronskian function $W[\psi_1(x),\psi_2(x)]=\psi_1(x) \psi_2^\prime(x)- \psi_1^\prime(x) \psi_2(x)=1$
which is known to be a constant of motion.
Here the prime denotes first differentiation with respect to $x$.
On the right, the RK-integration is carried up to (say) $x=w_b$ for the case of a finite range barrier $V_b$ in $V^F(x)$ (3). 
For infinite range cases like $V^{C}(x)$ (4) and $V^{D}(x)$ (5) RK-integration is to be carried up to (say) $x=D$ such that $V(D)$ 
is very small. Similarly, on the other side, the RK-integration is to be carried up to $x=-d-w_w$ in case of $V^F(x)$. In case of $V^{C}(x)$ (4)
and $V^{D}(x)$ (5) we integrate up to (say) $x=-D$. Let us denote the end values $\psi_1(-d-w_w), \psi_2(-d-w_w), \psi_1^\prime(-d-w_w),
\psi_2^\prime(-d-w_w)$ as $\psi_1, \psi_2,\psi_1^\prime,\psi_2^\prime$, respectively. The end values $\psi_1(w_b), \psi_2(w_b), \psi_1^\prime(w_b),
\psi_b^\prime(w_b)$ are denoted as $\phi_1, \phi_2, \phi_1^\prime, \phi_2^\prime$, respectively.

As RK-integration is step by step method wherein the calculated value of the function, $\psi(x)$,  and its slope (momentum) 
$\psi^\prime(x)$ at one step serve as initial values for the next step. This suits quantal calculations wherein the wavefunction 
and its derivative must match everywhere in the domain of the potential. Importantly, then it does not matter whether or not the 
potential is continuous or has a finite jump discontinuity at one or more number of points in the domain of the potential. 
We finally write the solution of Eq.~(6) as
\begin{eqnarray}
&&\psi(x)=A e^{ikx}+ B e^{-ikx},\quad -\infty < x \le -d-w_w \nonumber \\
&&\psi(x)=C_1 \psi_1(x)+ C_2 \psi_2(x),\quad  -d-w_w < x \le w_b \nonumber \\
&&\psi(x)=F e^{ikx},\quad x > w_b
\end{eqnarray}  
In case of $V^{C}(x)$ (4) and $V^{D}(x)$ (5), the distances $-d-w_w$ and $w_b$ will be replaced by $-D$ and $D$, respectively.
Next by matching $\psi(x)$ and ${d\psi(x) \over dx}$ at these points we get
\begin{eqnarray}
&&A e^{-ik(d+w_w)}+ B e^{ik(d+w_w)}= C_1 \psi_1 + C_2 \psi_2 \nonumber \\
&&ik(A e^{-ik(d+w_w)}- B e^{ik(d+w_w)})= C_1 \psi_1^\prime + C_2 \psi_2^\prime \nonumber \\
&&C_1 \phi_1 + C_2\phi_2= F e^{ikw_b} \nonumber \\
&&C_1 \phi_1^\prime + C_2 \phi_2^\prime = ik F e^{ikw_b}.
\end{eqnarray}
Solving Eqs.~(15), we get
\begin{eqnarray}
&&{B \over A}=-e^{-2ik(d+w_w)} {(\phi_1^\prime-ik \phi_1)(\psi_2^\prime-ik \psi_2)-(\phi_2^\prime-ik\phi_2)(\psi_1^\prime-ik \psi_1)
\over (\phi_1^\prime-ik \phi_1)(\psi_2^\prime+ik\psi_2)-(\phi_2^\prime-ik\phi_2)(\psi_1^\prime+ik\psi_1)},\\ \nonumber
&&{F\over A}=-{2ik e^{-2ik(d+w_w+w_b)} \over (\phi_1^\prime -ik \phi_1) (\psi_2^\prime + ik \psi_2)-(\phi_2^\prime-ik\phi_2)(\psi_1^\prime + ik \psi_1)}. 
\end{eqnarray}
Here we have used the constancy of the Wronskian $[\phi_1\phi_2^\prime-\phi_1^\prime \phi_2]=W[\phi_1,\phi_2]=1$.
The transmitivity (transmission probability) of the total the NWAFB is given by $T(E)$ as in above equation. This may be denoted fully as
\begin{equation}
T(E)=T(v_w, v_b, w_w, w_b, d, E), \quad T_b(E)=T(v_w=0, v_b, w_w, w_b, d, E),
\end{equation}
where $T_b(E)$ denotes the transmitivity of the (unperturbed) barrier and $v_w,w_w$ and $d$ may be taken to act as perturbation parameters.  

\section{Results and discussions}
Using the Eq.~(16), we calculate the transmitivity of various analytically intractable models given in section III.
Let us discuss the NWAFB represented by $V^F(x)$ in Eq.~(3).  
Figs.~3 presents $T(E)$ and $T_b(E)$ when $V_w(x)$ is a rectangular well in $V^F(x)$ (see dotted well in Figs.~1(b)). 
The form of the barrier is fixed as $V_b(x)=v_b xe^{-x^2}$ and its parameter $v_b=11.5$ this gives $(v_m)$ as 
about 5 units. In Figs.~3(a), we see only marginal excursions in $T(E)$ when the well is shallow, wide and distant. When the well is 
deeper but juxtaposed to the barrier ($d=0$) the frequency of oscillations decreases (see Figs.~3(b)). When the well is away from the barrier,
$T(E)$ is more oscillatory compare Figs.~3(b) with Figs.~3(c). When the depth of the well is increased to 10 units ($v_w > v_m$) the amplitude
of the  oscillations increases (see Figs.~3(d)). In NWABF the essence is that the oscillations in $T(E)$ are seen riding around $T_b(E)$. In 
other words the well induces oscillations in the transmitivity of the adjacent barrier. We would like to remark that a piecewise constant potential 
mentioned in Ref.\cite{8} (see Eq.~ (22) there) can now be seen as a NWAFB of the type (3), wherein both the well and the barrier are square (rectangular)
and $T(E)$ is oscillatory (see Fig.~5 there).

Next we study parabolic well in $V^F(x)$ (3). In Figs.~4(a), this time we find that the well-depth has to be comparable to the barrier height of 5 units
for changing  $T(E)$ appreciably when compared to $T_b(E)$. The effect of increase in the depth of the well can be seen to enhance the amplitude of 
of oscillations in $T(E)$ by comparing Figs.~4(a) with Figs.~4(c). $T(E)$ in Figs.~ 4(b) is less oscillatory as compared to that in Figs.~4(c) because 
the well and barrier are juxtaposed to each other with $d=0$. So in this model too the energy oscillations occurring in $T(E)$ are  due to 
increase in the width or depth of the well or its distance from the barrier. However, these oscillations are less prominent than those of rectangular 
potential model seen in Figs.~3. The general feature of the NWABF of the type $V^F(x)$ (3) that the transmitivity is more oscillatory when a thinner 
barrier is away from the well is well demonstrated when one compares Figs.~4(c) and Figs.~4(d).

The oscillations in the transmitivity of rectangular and parabolic models of NWAFB (3) may be attributed \cite{7} to their finite range
(finite support) and also to the distance $d$ over which the potential being zero allows the interference of plane waves. 
Further, the prominence of oscillations in $T(E)$ of rectangular model lies in the fact that rectangular 
potential well or barriers are most localized  profiles between two points than any other profile of finite support\cite{9}.

Fig.~5, displays the qualitatively similar oscillatory transmitivity when quite thin wells ($w_w=0.4$) are used in NWAFB of the
type given by $V^F(x)$ in Eq.~(3). The depths of the wells and their distances from the barrier are fixed as $v_w=10$ and $w_w=5$, respectively. 
These wells taken here are rectangular, parabolic, Gaussian, and triangular (see the line below Eq.~(3)). 
From this Fig.~5 we conclude that quite thin wells despite being away from the barrier
can induce prominent oscillations in $T(E)$ provided they are sufficiently deep. If not so deep the amplitude of oscillations will be small.

Now we study two more modifications of NWAFB which are made up of scattering potentials of infinite range. These are $V^{C}(x)$ (4)
and $V^{D}(x)$ (5). In the case of $V^{C}(x)$ (see solid curve in Figs.~1(c)) when the well and the barrier are juxtaposed continuously at $x=0$,
we find (see Figs.~6(a)) that if the well is strong it reduces the transmitivity and then increases it only marginally at energies below the barrier. 
At energies above the barrier height the changes are inappreciable. 
In the dis-continuous case (see dashed curve in Figs.~1(c)), we find that the hidden well reduces $T(E)$ over all (below and above the
barrier) energies (see Figs.~6(b)). This is the characteristic feature of the potential being discontinuous at a point ($x=0$) as the well and the barrier
are juxtaposed there in a discontinuous way as in the case of a simple potential step\cite{2,3,4}. Also the well reduces transmitivity of the barrier 
in an appreciable way only if it is strong (e.g., $w_w >w_b$). We have confirmed absence of energy oscillations in these two models by varying $v_w$ and $v_b$
high and low abundantly.
Moreover, in this regard the exact analytic expression \cite{8} $T(E)$ of the Scarf II potential ($V(x)=V_0 \tanh x ~{\rm sech} x$) readily testifies to a non-oscillatory 
behaviour of NWAFB of the type (4) as a function of energy
\begin{equation}
T(E) =\frac {\sinh^2 2\pi k}{[(\cosh 2\pi k +\cos 2 \pi p) (\cosh 2\pi k+ \cosh 2\pi q)]},
\end{equation}
with $k=\sqrt{E/\Delta}$, $p=\mbox{Re}\,(\sqrt{1/4+iV_0/\Delta})$,
$q=\mbox{Im}\,(\sqrt{1/4+iV_0/\Delta})$, and $\Delta= \hbar^2/(2ma^2)$.

However, in the above models $V^{C}(x)$ (4) and $V^{D}(x)$ (5) if the well and barrier are separated by a distance, $d$,
the transmitivity will again acquire oscillations. We would like to emphasize that it is the separation between the
well and the barrier that plays a crucial role in causing energy-excursions (oscillations) in $T(E)$ with respect to $T_b(E)$.

Figs.~6(c,d) demonstrate that in case of single piece NWAFB (1) when $v_b=5$ and $d=8$ it requires a very deep well
$(v_w=2000)$ to get even small excursions in $T(E)$ with respect to $T_b(E)$. Appreciable energy oscillations can be seen in
$T(E)$ only if the well is much deeper ($v_w=5000$). This feature is surprising in view of the fact that the NWAFB
of the types (Eqs.~(2,3)) in Figs.~2-5 have displayed good energy oscillations even if $v_w$ is twice of $v_b$ or even 
less than $v_b$.

In all the results presented in Figs.~2-6 (see the dotted curve), in NWABF the general trend of $T(E)$ is determined by the barrier is
irrespective of the strength of the well. Broadly, three (Eqs.~(1-3)) types of NWAFB (see Figs.~ 1) entailing single well and a 
single barrier are possible. However, one has choices of the profiles for the well and the barrier in them. Apart from the results 
of various profiles presented here in Figs.~(2-6) we have also studied several other profiles and explored  various parametric regimes 
in all three types of NWABF to confirm our findings presented here. 

\section{Conclusions} 
The transmission through a barrier is the phenomenon of positive energy continuum, we conclude that the well (at negative energies)
essentially causes energy-excursions (ripples or oscillations) in the transmitivity of the barrier. Howsoever strong the well is 
the trend of transmitivity as a function of energy is determined only by the barrier.
Ordinarily, the finite support(range)
of the well may also be attributed\cite{7} to cause  energy oscillations in the 
transmitivity. In this regard, the energy-oscillations in the transmitivity of one-piece smooth potential (1) of infinite range found 
here are unexpected. However, it has required the well depth to be extremely large (see Figs.~6(d)). 
The separation between the well and the barrier is {\it sufficient} if not the {\it necessary} condition in giving rise to 
oscillations in transmitivity. When the well and the barrier are separated away, the potential in the intermediate region is zero.
This gives a scope for destructive and constructive interference of plane waves and hence the frequency of energy-oscillations
in the transmitivity increases. However, if one views the well as a perturbation to the barrier then the enhanced oscillations
in $T(E)$ despite the well being distant is paradoxical. The infinite range well and barrier if joined at a point with no separation 
($d=0$) between them do not seem to have energy-oscillations in transmitivity until they are separated.

The energy-oscillations in transmitivity at energy below the barrier suggests 
a novelty because usually transmitivity is found\cite{7,8,9,10} to be oscillatory at energies above the barrier. 

The transmitivity of various potential systems which converge asymptotiacally $(x \rightarrow \pm \infty)$ to zero or to a constant value and which
are either continuous or entail finite jump discontinuities can be found using Eq. (16) presented here.
In this article we have presented the first and hopefully an exhaustive study of transmission through non-overlapping well adjacent 
to a finite barrier. We hope that this investigation will be found pedagogically valuable.  

\section{Appendix}
\appendix{}
\numberwithin{equation}{section}
\section{}

The Runge-Kutta\cite{6} solution of the coupled first order equations
\begin{equation}
y^\prime=f(x,y,z), z^\prime=g(x,y,z),
\end{equation}
are obtained as $y_1,y_2,y_3,...,y_n$ and $z_1,z_2,z_3,...z_n$ starting with the initial values $y_0,z_0$ using the following equations.
\begin{eqnarray}
&&y_{n+1}=y_n+{h \over 6} [k_1+2k_2+2k_3+k_4], \quad z_{n+1}=z_n+{h \over 6}[m_1+2m_2+2m_3+m_4],~ n \ge 0,~h={D \over n}\nonumber \\
&&k_1=f(x_n,y_n,z_n), \quad m_1=g(x_n,y_n,z_n)\nonumber \\
&&k_2=f(x_n+h/2,y_n+h k_1/2,z_n+h k_1/2), \quad m_2=g(x_n+h/2,y_n+h m_1/2,z_n+h m_1/2) \nonumber \\
&&k_3=f(x_n+h/2,y_n+h k_2/2,x_n+h k_2/2), \quad m_3=g(x_n+h/2,y_n+h m_2/2,x_n+h m_2/2) \nonumber  \\
&&k_4=f(x_n+h,y_n+h k_3, z_n+h k_3), \quad m_4=g(x_n+h,y_n+ h m_3, z_n+h m_3).
\end{eqnarray}
When we solve (11) for $y_0=1,z_0=0$, we get $\psi_1(x)$ and $\psi_1^\prime(x)$ and we get $\psi_2(x)$ and 
$\psi_2^\prime(x)$ when the starting values are $y_0=1,z_0=0$.

\begin{figure}
\centering
\includegraphics[scale=1.9]{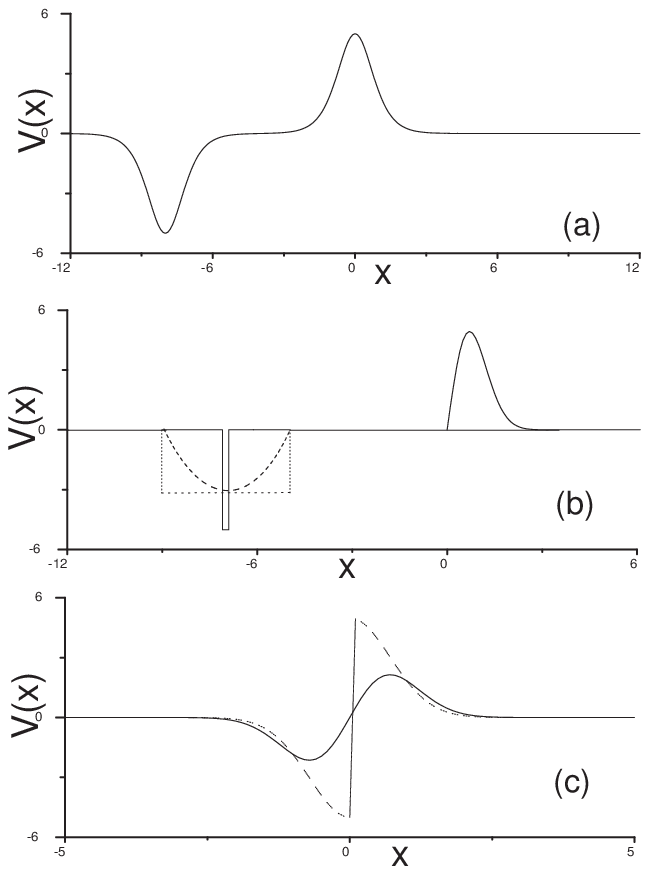}
\caption{The schematic depiction of various NWAFB. (a): single piece smooth potential (1), (b) $V^F(x)$ (3): parabolic well
(dashed line), rectangular well (dotted line) and  very thin rectangular well near a barrier, (c) $V^C(x)$ (4): a smooth well 
continuously juxtaposed to a barrier (solid line) and $V^D(x)$ (5): a smooth well discontinuously juxtaposed to the a barrier.}
\end{figure}

\begin{figure}
\centering
\includegraphics[scale=2]{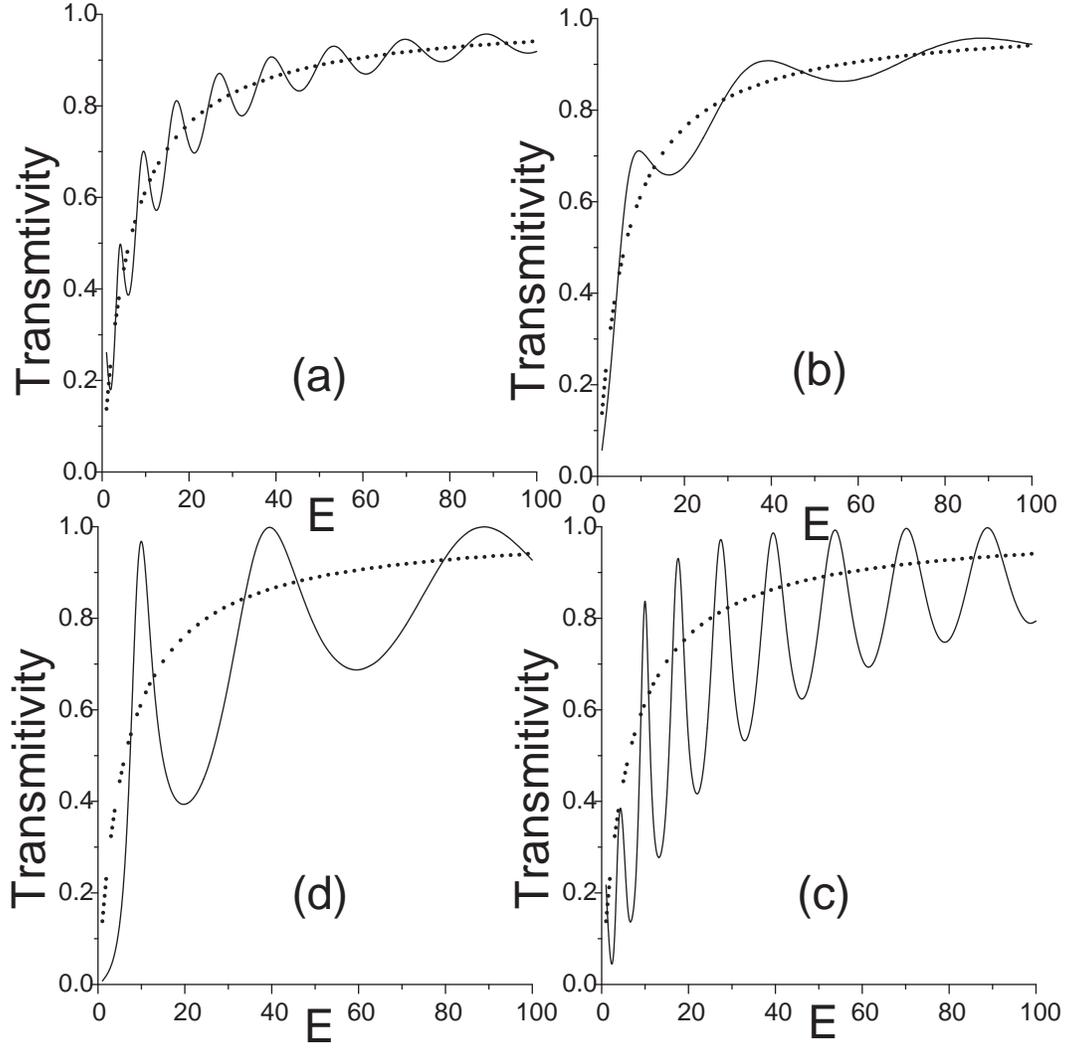}
\caption{The solid line represents the transmitivity, $T(E)$, of the delta potential model $(V^{\delta})$ of NWAFB  (2). The dotted curve represent
the transmitivity, $T_b(E)$, of the barrier only. We have a fixed barrier height $V_b=5$ and take(a): $v_w=1,d=3$, 
(b): $v_w=1, d=1$, (c) $v_w=5,d=3$, (d) $v_w=5, d=1$.}
\end{figure}

\begin{figure}
\centering
\includegraphics[scale=2]{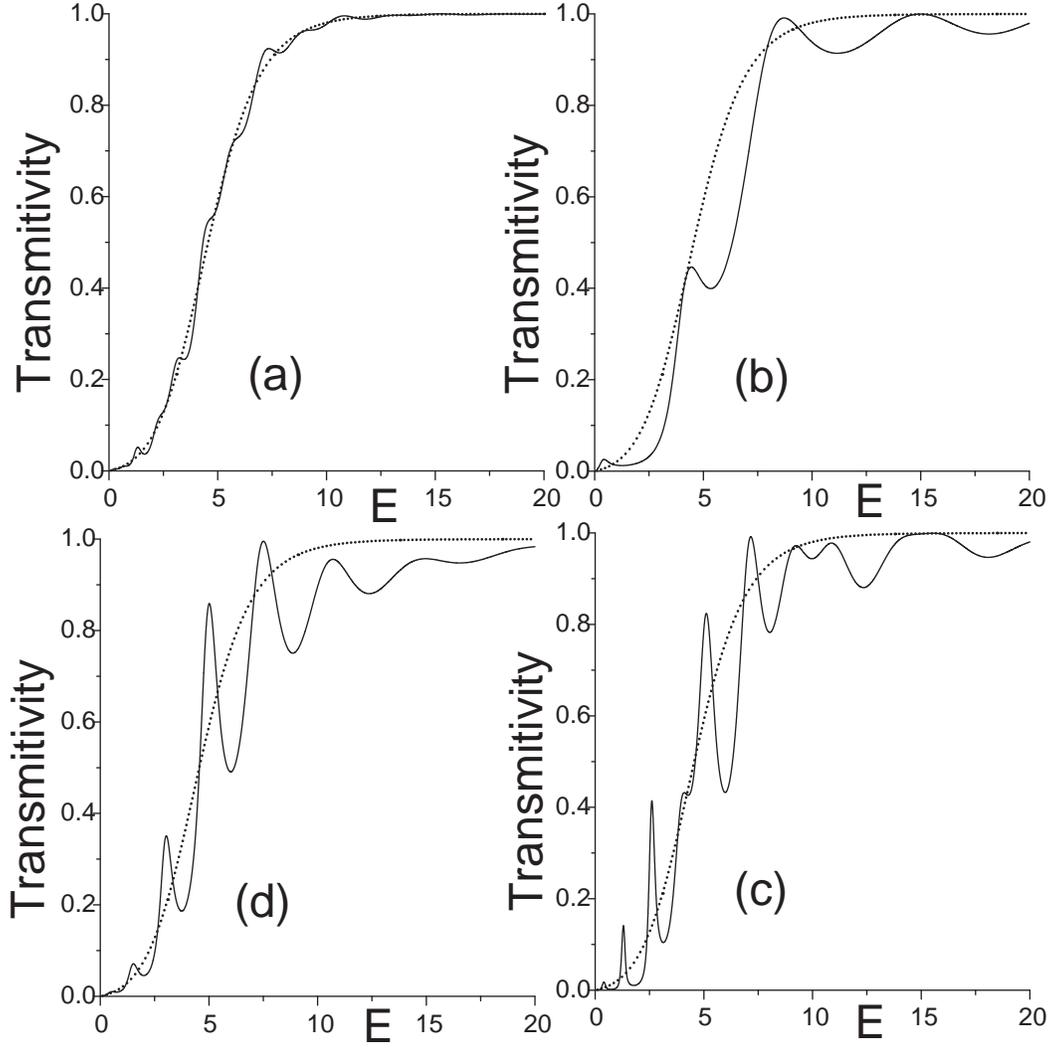}
\caption{The same as in Figs.~2 for the NWAFB of the type $V^F(x)$ (3). Here the barrier  $V_b(x)=v_b x e^{-x^2}, v_b=11.5$
is perturbed by a rectangular (square) well. The effective height of the barrier $v_m$  is approximately 5 units.
We have taken (a): $v_w=1, d=5, w_w=5$, (b): $v_w=10, d=0, w_w=5$, (c): $v_w=10, d=5, w_w=5$, (d): $v_w=10, d=5, w_w=1$.} 
\end{figure}

\begin{figure}
\centering
\includegraphics[scale=2]{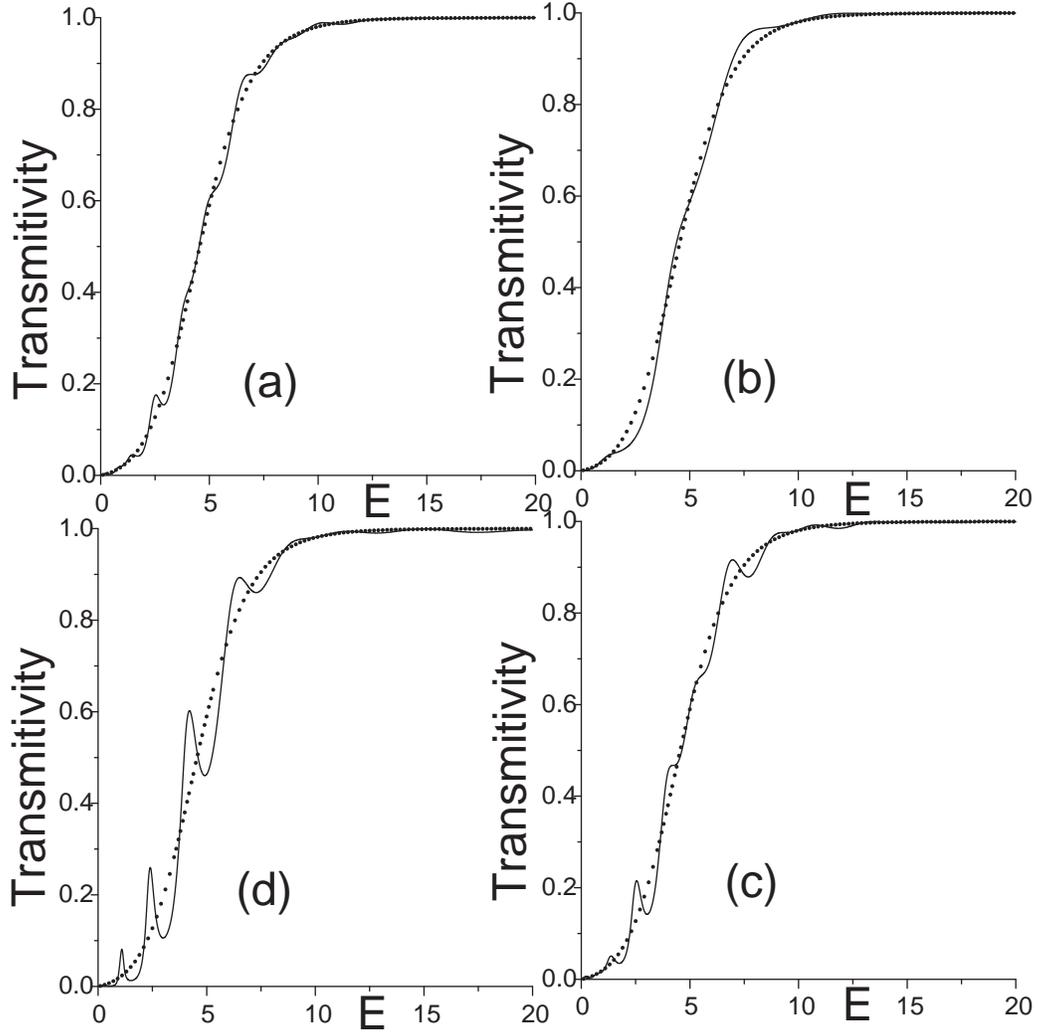}
\caption{The same as in Figs.~2 for the  NWAFB of the type $V^F(x)$ (3). Here in general the energy-oscillations 
in $T(E)$ are present but these are less prominent than those in Figs.~2.  The same barrier($V_b$) is now perturbed by a parabolic
well of finite range. We take (a): $v_w=5, d=5, w_w=5$, (b): $v_w=10, d=0, w_w=5$, (c): $v_w=10, d=5, w_w=5$, (d): $v_w=10, d=5, w_w=1$}
\end{figure}

\begin{figure}
\centering
\includegraphics[scale=1]{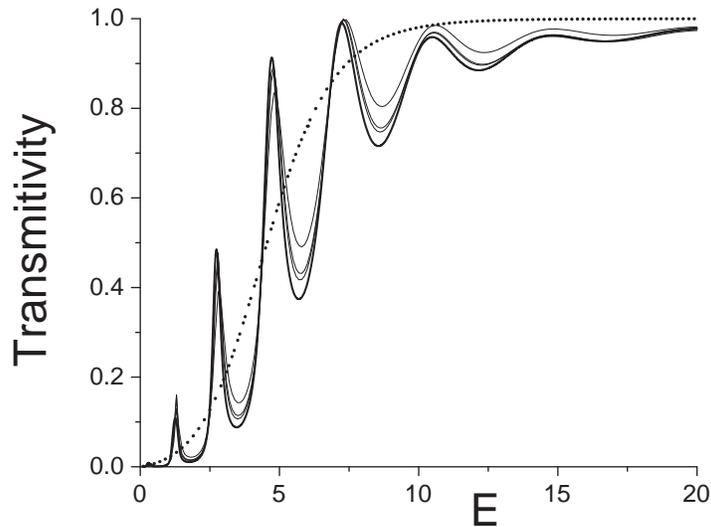}
\caption{$T(E)$ (solid lines) and $T_b(E)$ (dotted curve) for various NWAFB of the type $V^F(x)$ (3) when the wells are quite thin($w_w=0.4$).
We have $v_w=10, d=5$. These wells are rectangular, parabolic, Gaussian, and triangular used in Eq.~(3) (see the text below Eq~(3)).
Thin wells away from the barrier give rise to qualitatively similar transmitivity which is oscillatory. 
This is an essential feature of the NWAFB of the type in Eqs. (2,3).}  
\end{figure}

\begin{figure}
\centering
\includegraphics[scale=2]{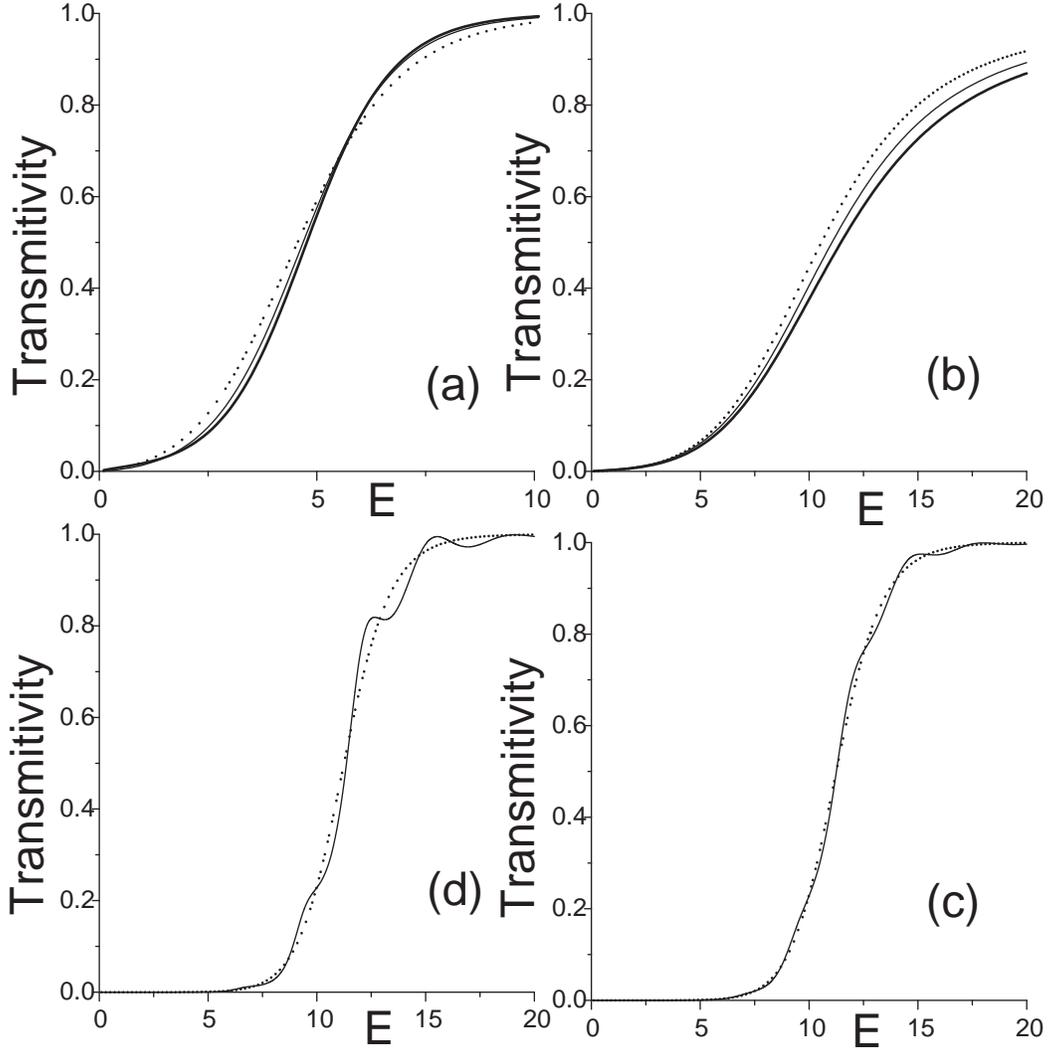}
\caption{Transmitivity, $T(E)$ for (a): the continuous (4) and (b): the discontinuous (5) models;
the dotted line $(v_w=5)$, thin solid line $(v_w=10)$ and thick solid line $(v_w=15)$.
Figs.~(c,d) represent the transmitivities for the single piece smooth NWAFB (1). 
For a fixed distance ($d=8$) between the well and the barrier and $v_b=5$ Figs.~(c) shows only small
excursions in $T(E)$ only when the well is very deep ($v_w=2000$). In Figs.~(d), significant oscillations in $T(E)$ have 
required even higher value $v_w(=5000$).}  
\end{figure}

\end{document}